# A Neural Network Approach for Improved Seismic Event Detection in the Groningen Gas Field, The Netherlands

Bob Paap[1*], Stefan Carpentier[1], Peter-Paul van Maanen[1] and Sjef Meekes[1]

[1]TNO, Utrecht, The Netherlands

*Corresponding author:
Bob Paap
TNO
Utrecht, The Netherlands
bob.paap@tno.nl

January 20, 2020

**Abstract**
Over the past decades, the Groningen Gas Field (GGF) has been increasingly faced by induced earthquakes resulting from gas production. The seismic monitoring network at Groningen has been recently densified to improve the seismic network performance, resulting in increasing amounts of seismic data. Although traditional automated event detection techniques generally are successful in detecting events from continuous data, its detection success is challenged in cases of lower signal-to-noise ratios. The data stream coming from these networks has initiated specific interest in neural networks for automated classification and interpretation. Here, we explore the feasibility of neural networks in detecting the occurrence of seismic events. For this purpose, a three-layered feedforward neural network was trained using public data of a seismic event in the GGF obtained from the Royal Netherlands Meteorological Institute (KNMI) data portal. The first arrival times and duration of earthquake waveforms determined by KNMI for a subset of the station data, were used to detect the arrival times and event duration for the other uninterpreted station data.

Subsequently various attributes were used as input for the neural network, that were based on different short term averaging/long term averaging (STA/LTA) and frequency sub-band settings. Using these input data, the network's parameters were iteratively improved to maximize its capability in successfully discriminating seismic events from noise and determine the event duration. Results show an increase of 65% in accurately detecting seismic events and determining their duration as compared to the reference method. This clears the way for improved interpretation of signal waveforms and automated seismic event classification in the Groningen area.

**Introduction**
The drivers behind the development of seismic event classification and interpretation capacity include the rise of broadband, digital and wireless seismic acquisition networks together with application of seismic processing techniques on significantly sized datasets while maintaining acceptable runtimes. The ongoing growth in amount and size of seismic networks results in vast amounts of continuous data-streams which in turn requires large computing facilities to allow data processing. Here we consider the field of seismicity monitoring where the challenge in processing the data-streams is two-fold: 1) in an initial step events have to be detected in realtime, preferably at the seismic network stations themselves. This requires a fast, efficient and reliable decision-making algorithm. Isolating these events from continuous data streams strongly reduces the amount of data transfer from seismic stations within a network to a central server. 2) In a second step subsequent seismic processing, event classification and characterization should take place.

Several seismic monitoring networks are deployed in the Netherlands to monitor natural and induced seismicity. They are required to safeguard conformance, production and safety regulations around oil, gas, geothermal and $CO_2$ exploitation sites. Special interest goes to the Groningen Gas Field (GGF), the tenth-largest gas field in the world. This field is faced by induced seismicity that is associated with gas production, with local magnitudes reported up to M=3.6. Therefore, the monitoring network at the GGF has been densified significantly over recent years, and nowadays consists of accelerometers and broadband seismometers positioned at the surface, and geophone strings placed in boreholes [*Dost et al.*, 2017]. The large data stream coming from the Groningen network makes it well-suited to explore the potential of (automated) classification and interpretation of seismic events. This requires data processing of continuous data-streams, which typically consists of data conditioning and preprocessing, followed by automated event detection and successive source characterization.

Currently, for the GGF seismic events contained in sensor recordings are detected in a three-step procedure: First, the onset of a possible seismic event in the network signal is detected by invoking an automated trigger, usually based on an amplitude threshold or a ratio of subsequent amplitudes. Second, an AR-AIC picker [*Sleeman and van Eck*, 1999] is applied to perform a more accurate arrival time picking. Third, a human interpreter inspects the detected event, confirms whether the event is genuine, compares the signal with those collected by other sensors and then determines event properties, such as magnitude and location.
Various techniques for automated event detection exist, such as the ratio of the STA/LTA picking in the time domain (i.e. energy transient [*Vanderkulk et al.*, 1965]), STA/LTA in the time-frequency domain (i.e. frequency transient [*Withers et al.*, 1998]), cross-correlation with template waveforms [*Gibbons and Ringdal*, 2006] and artificial neural network (ANN) approaches [e.g. *Curilem et al.*, 2009; *Doubravová et al.*, 2016; *Tiira*, 1999].

Although these techniques generally are successful in detecting events from continuous data, the detection reliability is challenged in case of lower signal-to-noise ratios (i.e. unknown amounts of missed events). This has initiated specific interest in the use of ANNs for automated classification of seismic events for evaluating and interpreting potential seismic events. There are various studies that successfully applied ANNs to detect seismic events [*Romeo*, 1994; *Dai and MacBeth*, 1995; *Gentili and Michelini*, 2006; *Curilem et al.*, 2009; *Doubravová et al.*, 2016; *Ross et al.*, 2018].

Here, we specifically explore the potential of ANNs for detecting induced earthquakes at the GGF. The expected power of the ANNs is that it can efficiently detect earthquake signal within continuous ground motion recordings, especially on recordings with a relatively low signal-to-noise ratio, while benefiting from faster computation times compared to conventional detection methods. If we recall the two steps of seismic data processing described before in the Introduction, the power of ANN's becomes even more apparent. ANN's could be more efficient in performing initial detection in step 1 due to faster computation times. This holds especially for sparsely layered ANN's having low computation times. In the second step conventional seismological processing, possibly in combination with additional ANN's, can be used to further characterize detected seismic events. In this study we focus our efforts on the first step: explore the potential of ANN for efficient and reliable event detection. To the best of our knowledge there is no study that specifically addresses the application of ANNs on detecting induced seismic events in the GGF.

For this purpose a simple feedforward ANN with three layers was employed for seismic event detection using the KNMI network at the Groningen site. When designing an ANN it is common practice to address supervised problems by using one hidden layer, provided there is a limited complexity of the problem and that the user-defined input attributes are meaningful for the considered dataset. Seismic recordings from the extensive Groningen monitoring network were collected for three different induced earthquake events, and seismic traces were labeled by a seismologist. Different combinations of attributes of the seismic trace were used as input for the ANN. The performances of the ANNs were evaluated and compared, taking into account both their sensitivity and specificity.

**Method**
In this section we first discuss the characteristics of the selected dataset, as well as the procedure for defining the target function for the ANN by labeling the data. Next, we present the details of the reference event detection method (STA/LTA and duration), followed by the neural network event detection method. Finally, we explain the analyses that will be used to compare the results obtained with the different detection methods against the target function.

*Dataset*
In this study we used public earthquake data obtained from the KNMI data portal (see Data and Resources Section). To challenge the ANN, we selected station data of the Groningen network from three different earthquakes induced by gas exploitation having different magnitudes and epicenters and occurred at approximately 3 km depth. The three events are termed Siddenburen, Wagenborgen and Garsthuizen, respectively having local magnitudes of 0.8, 1.2 and 2.8 (see Figure 1). All available vertical component data for the considered earthquakes was used and we did not restrict ourselves to recordings with a relatively high signal-to-noise ratio. In total this consisted of 832 vertical receiver recordings that were recorded by surface and borehole sensors (see Figure 1). Borehole sensors are placed in 70 boreholes equipped with geophones vertically spaced 50 m up to 200 m depth. The data were downloaded as mini-SEED files from the KNMI webportal having 60 s before and 120 s after origin time.

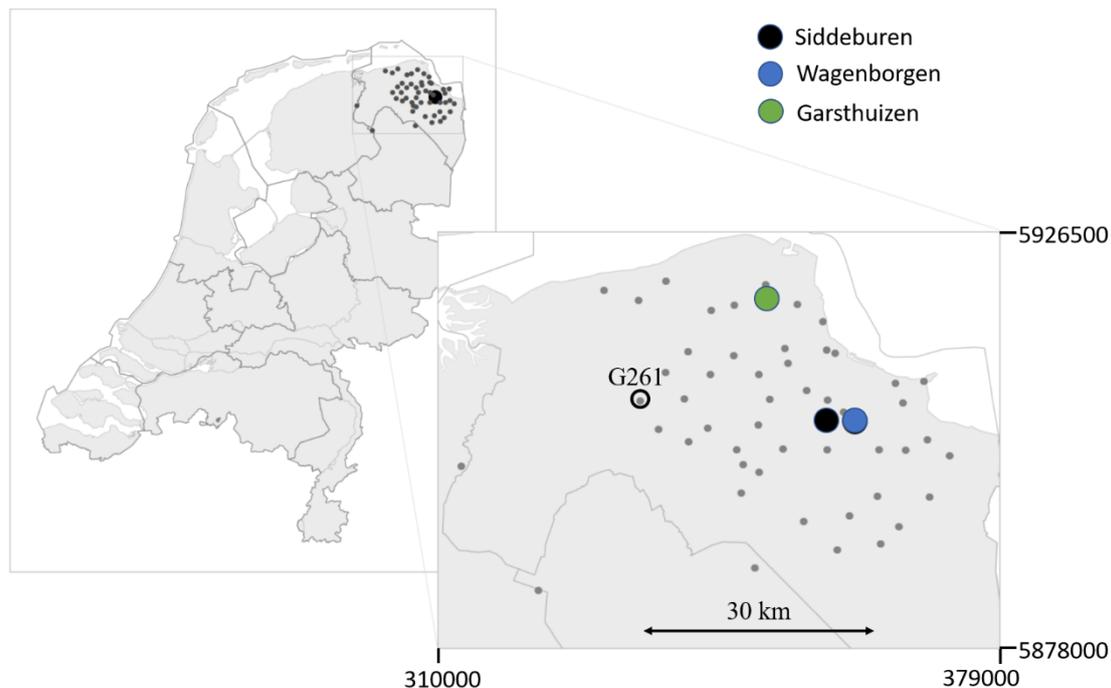

**Figure 1.** Map of all available vertical receivers of the three earthquakes considered in this study 21st 2016. The earthquake epicenters are marked by the filled colored circle. Station G261 is highlighted for reference in the Results Section and marked by the open circle. The coordinates are specified in Universal Time Mercator in meters.

Before labeling the seismic data, standard seismic pre-processing was conducted with Matlab using the mseed-package [*Beauducel*, 2012]. A high pass frequency filter was applied with a low-cut frequency of 3Hz to reject low-frequency noise, as well as a notch filter at 50Hz to reject spurious signals. To train and test the ANN, a target function needs to be defined against which the result of the ANN can be compared. This target function will be used as well to evaluate the performance of the reference STA/LTA detection method (discussed in the next section). The target function requires labeling of the seismic data, which is done in a supervised approach. The data is labeled either as having the absence (label 0) or presence (label 1) of an earthquake event signal. The publicly accessible catalogue of KNMI only contained the labeled presence of events (i.e. label 1), but lacked the labeled absence of events (i.e. label 0). Therefore, we extended the labeling procedure by labeling non-events within the raw continuous data. In this way with our approach we can identify both 'true positives' (i.e. correctly labeled events) and true negatives (i.e. correctly labeled noise /non-events). To further illustrate this labeling procedure we now consider the Wagenborgen event, for which the KNMI catalogue contains precisely determined p-wave arrival times for 29 out of a total of 204 vertical receiver traces. These traces are plotted against offset in Figure 2b. We determined the first arrival times for all 204 traces from a linear regression fit through the scatter points in Figure 2b, resulting in a velocity of 4860ms$^{-1}$ ($v_{app;start}$) and a $t_{0,start}$ of 1 s (marked by upper black dashed line Figure 2)a. Hence the time samples within the remaining data (i.e. 204 - 29 = 175 traces) was identified as 'event' for times larger than $t_{0,start}$ + offset/$v_{app,start}$. For the end-time of the event an apparent velocity of 1176ms$^{-1}$ ($v_{app,end}$) was defined, highlighted by the lower dashed black line in Figure 2a. Data recorded after the interpreted end of the event were not used for training the neural network. Note that in our approach we train the neural network on individual time samples, which is favored for our purpose of accurately detecting both p-wave onset time and event termination time. This approach implies that each time sample within a seismic trace is labeled as either ' event' or 'no event', depending whether part of an earthquake waveform is present at that respective time or not. This amounted to a total of approximately 1.6 million time samples for all 832 seismic vertical receiver recordings of the three considered earthquakes combined.

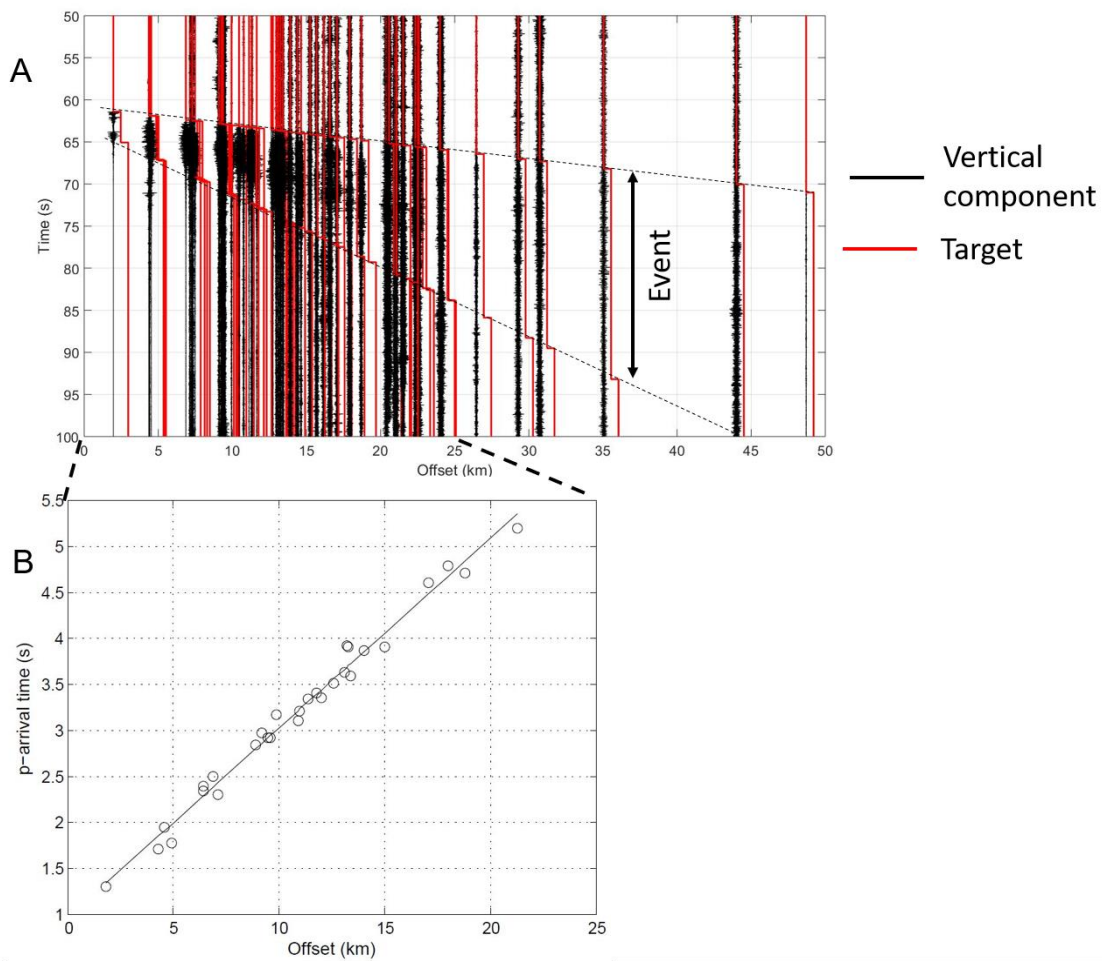

**Figure 2.** a) Waveforms of the Wagenborgen event recorded at vertical components as a function of offset. In red the result of the manual interpretation of the onset and termination of an event is plotted. b) Linear fit through p-wave arrival times for a selected range of offsets from the Wagenborgen event. This selected offset is only shown to enhance visualization of the distribution of scatter points, and during training waveforms for all recorded offsets were used.

*STA/LTA detection method and duration (reference method)*
As STA/LTA detectors are widely used and are the primary means for initial detection of seismic events [*Gentili and Michelini*, 2006], the STA/LTA method as used by the KNMI for the Groningen area was adopted and applied on the data with a short time window of 0:5 s, a long time window of 10 s, a trigger on threshold of 3:0 and a trigger off threshold of 1:5. We adopted the processing flow and parameters settings as used by KNMI for the Groningen data. This consisted of running a mean high-pass filter with a time window of 10 s, a one-sided cosine taper with a time window of 30 s and a bandpass filter with 4 corners, a minimum frequency of 5Hz and a maximum frequency of 40Hz. The event duration was combined with STA/LTA to make a fair comparison with the performance of NN, as both yield a binary classification of 'event' or 'no event' for each individual time sample along the seismic trace, respectively 1 or 0. The STA/LTA method on the other hand only identifies an event over a time-window much shorter than the actual extent of the full waveform. STA/LTA cannot discriminate increased amplitudes in the coda of the waveform, because of the gradual attenuation of the signal resulting in STA/LTA falling below the threshold value. Different definitions and relationships exist to estimate the duration of an event described in *Bommer et al.* [2009], each having its pro's and con's. Here, we adopt the significant duration, which is a measure of the duration of an earthquake signal and described by *Bommer and Martínez-Pereira* [1999]; *Bommer et al.* [2009]. The time window corresponding to the significant duration starts at the onset time, which is provided by the outcome of the STA/LTA algorithm. Next, the significant duration, which is an unscaled expression of the Arias intensity, is calculated by:

$$AI_{unscaled} = \int_0^{tr} a^2(t)dt$$

Where a(t) is the acceleration time series and tr is the total length of the seismic event.
This equation is similar to the definition of the Arias Intensity (equation (1) in *Bommer et al.* [2009]), except that we neglect the gravitational constant which is only used for scaling purposes. We define the significant duration by the interval in which 5% to 95% of the total integral is accumulated. To ensure that the temporal coverage of the event is sufficient, we impose that the time window used for duration estimation starts 5 s before and 10 s after the STA/LTA trigger on and off, respectively.

*Neural network detection method*
Typically when using neural networks with one hidden layer, the hidden layer has a number of neurons being approximately the mean of the number of neurons in the input and output layers. Similarly, we defined 14 neurons in the input layer, 6 in the hidden layer and 1 in the output layer. The neurons in the input layer have identity activation functions, while the neurons in all other layers have sigmoid activation functions. The seismic data from which attributes were calculated were not processed and remained in raw condition, such that the ANN has maximum freedom to decide how to use the data attributes for training the neural network.

We defined input attributes based on temporal amplitude variations using different STA/LTA ratios, together with different frequency density subband attributes. Both attribute types are known to be informative indicators for the presence of coherent seismic energy. Based on visual inspection we iteratively optimized the time windows and thresholds for STA/LTA, as well as the discretization of frequency-density subbands attributes, such that attributes contained complementary information about the characteristics of the seismic event. This resulted in 4 different STA/LTA window settings and frequency density calculated in 10 successive frequency subbands, all verified to contain complementary information regarding the onset of a seismic event.

Figure 3 shows the standard three-layer feedforward neural network that describes the different neural networks that were used in this study. As is shown in this figure, the neural networks have a maximum of 14 different input attributes. Three networks (named NN1, NN2 and NN12) were implemented and tested using different combinations of these input atributes:

- NN1 consisting of 4 input attributes: 4 STA/LTA ratios of varying short and long term windows as input (0.5 s/10 s , 5 s/15 s , 5 s/30 s , 10 s/30 s )
- NN2 consisting of 10 input attributes: 10 mean power spectral densities of sequential frequency intervals between 0Hz to 50Hz using short-time Fourier transform.
- NN12 consisting of 14 input attributes: all of the above

Based on the 'no event' and 'event' labeling of the individual time samples in the 200 seismic data, the neural networks were trained using the Newton conjugate-gradient algorithm [*Hestenes and Stiefel*, 1952]. The trained neural networks can be used to automatically label data in other seismic records either as having an absence ('no event') or presence ('event') of earthquake signal. Furthermore the ANN provides information on the precise onset and termination times of the earthquake waveform from the individually labeled time samples.

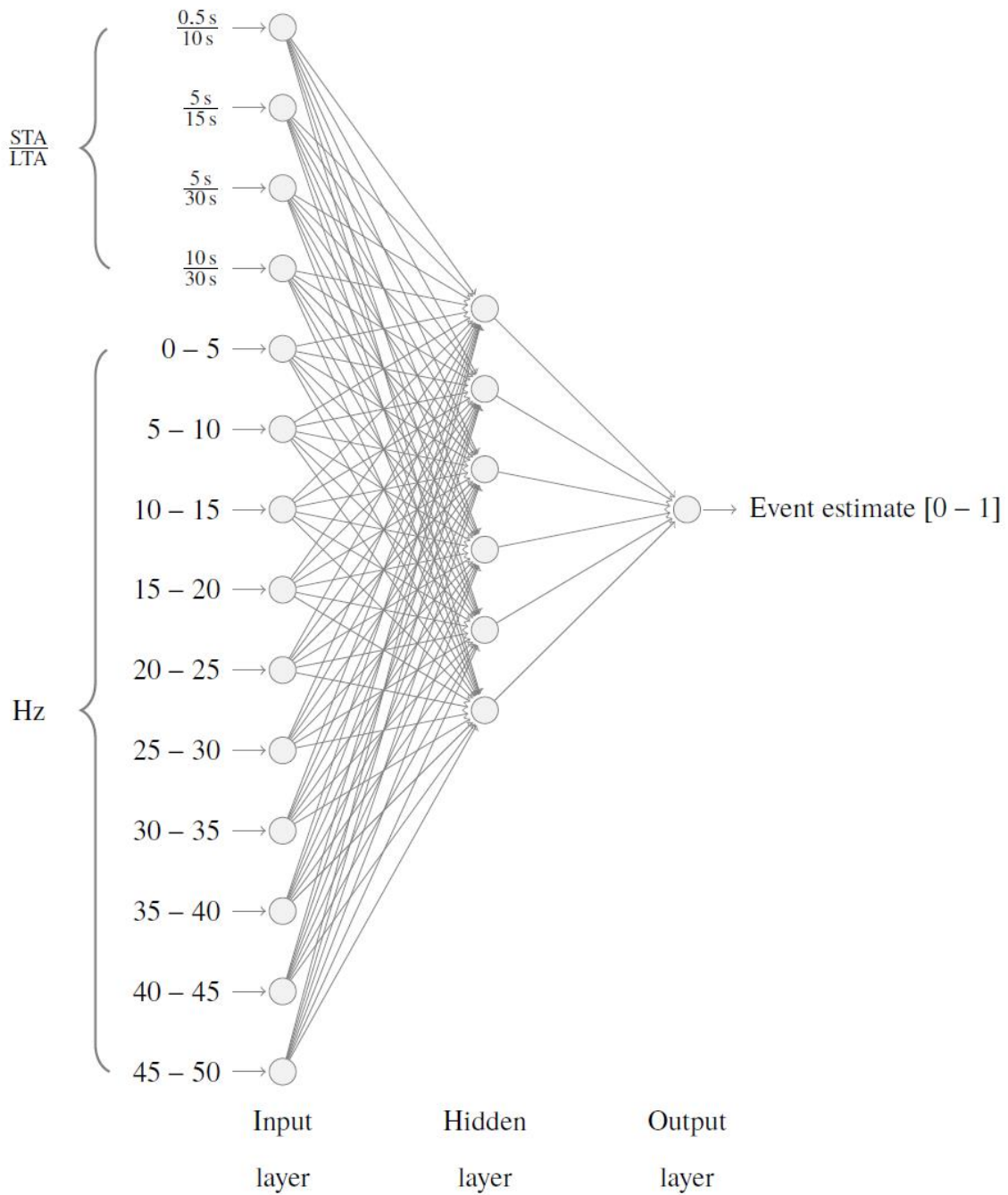

**Figure 3.** The three-layer feedforward neural network used in this study.

*Analyses*

To calculate the performance of the detection methods STA/LTA, NN1, NN2, and NN12, stratified k-fold cross-validation was applied to the results *Stone* [1974]. This means that for each detection method 20 independent and stratified folds (i.e. containing 19/20 of the data with an equal balance of events vs. no events) a neural network was trained and successively applied on the remaining 1/20 of the data. For each of the 20 testing results the performance was measured by calculating the sensitivity index:

$$d'=\text{true positive rate-false positive rate}$$

Statistical analysis was done using a repeated measures ANOVA to indicate any effect on detection performance for the given methods (by means of a statistical F-test [*Vonesh and Chinchilli*, 1997]). Then post-hoc dependent t-tests [*Jaccard et al.*, 1984] were performed to test the individual hypotheses that indeed the mean d0s of the neural networks is higher than of STA/LTA. A Receiver Operator Characteristic (ROC) curve [*Swets and Pickett*, 1986; *Woods and Bowyer*, 1997] of the best performing neural network detection method is constructed to show its true positive rate versus false positive rate trade-off: Higher true positive rates of the NN method may result in acceptable higher false positive rates and lower false positive rates may result in acceptable lower true positive rates as compared to the STA/LTA method. From the ROC curve the mean d0 of the NN method is then compared against that of the STA/LTA method for equal false positive rates. Increased performance of ANN does not necessarily mean higher false positive rates as compared to STA/LTA. To further evaluate the performance of neural networks, a comparison was made between the results of the neural network trained and applied on one earthquake (only the Wagenborgen event) against training and application on three earthquakes (i.e. Siddeburen, Wagenborgen and Garsthuizen).

**Results**

To illustrate the outcomes of the different detection methods, Figure 4 shows an example of the outcomes of the NN12 method for the Garsthuizen event recorded on the vertical component of station G261. The station location is indicated in Figure 1. Figure 4 shows both true positive and false positive outcomes for both the conventional and the neural network detection approach. The left column (a) shows the outcome of the neural network based on all 14 input data derivatives (NN12). Here, 20 realizations of the classification are shown for each time sample along the waveform ranging from 0 to 1. Here, we defined a threshold of 0.5, above which a sample is classified as an event, which are respectively highlighted by the horizontal gray lines. In (b) the target function is shown based on the manual interpretation of the event. The result of STA/LTA detection method combined with the event duration is shown in (c). The initial seismic trace is shown in (d). The input set of 10 frequency subbands between 0 and 50 Hz are shown in (e), and the input set of 4 different STA/LTA bands are shown in (f). In Figure 4 the neural network outcome in (a) shows a close resemblance in onset time and event duration with the target (b) from the onset time at 34 s, until the end of the event. The onset time of the event found by STA/LTA (c) is delayed by 3 s and its corresponding duration (d) is 6 s shorter compared to the target function. This example shows that the neural network is capable of correctly identifying event samples in the tail of the event where signal-to-noise ratios gradually decrease. Additional differences between the two detection methods are observed when considering pre-event false-positive detections. Although both neural network and STA/LTA outcomes contain false positives, on a sample basis STA/LTA significantly outnumbers the neural network in false positives. The neural network outcome in Figure 4 does contain some false positive samples prior to the actual start of the seismic event (0 s to 34 s), which are represented by thin horizontal gray lines in Figure 4a. The STA/LTA and duration result shows a full window from 3 s to 10 s with false positives (continuous gray band), which amounts to approximately 1400 false positive time samples based on 200Hz sampling frequency. This illustrates that the amount of false positive samples for the STA/LTA performance dramatically increase. This results from the definition of the significant duration, which translates each exceedance of STA/LTA into an event with a minimum duration of 15 s, assigning false positive values to all time samples contained in this time window.

In Figure 5 the performances of the different detection methods are shown for the neural network trained both on 1 event (Figure 5a) and on three events (Figure 5b). A further explanation on the differences in neural network performance between one and three events is given in the discussion. For the neural network trained on 3 events (Figure 5b), a repeated measures ANOVA shows a significant effect of detection method on performance, $F(4,16) = 1184:21; p < :0001$. Post-hoc dependent t-tests show significant improvements for the detection methods as compared from left to right in Figure 5 (all $p = 0:00$). This means that supplying more input attributes (when going from NN1 to NN2 to NN12) leads to better performance of the classifier.

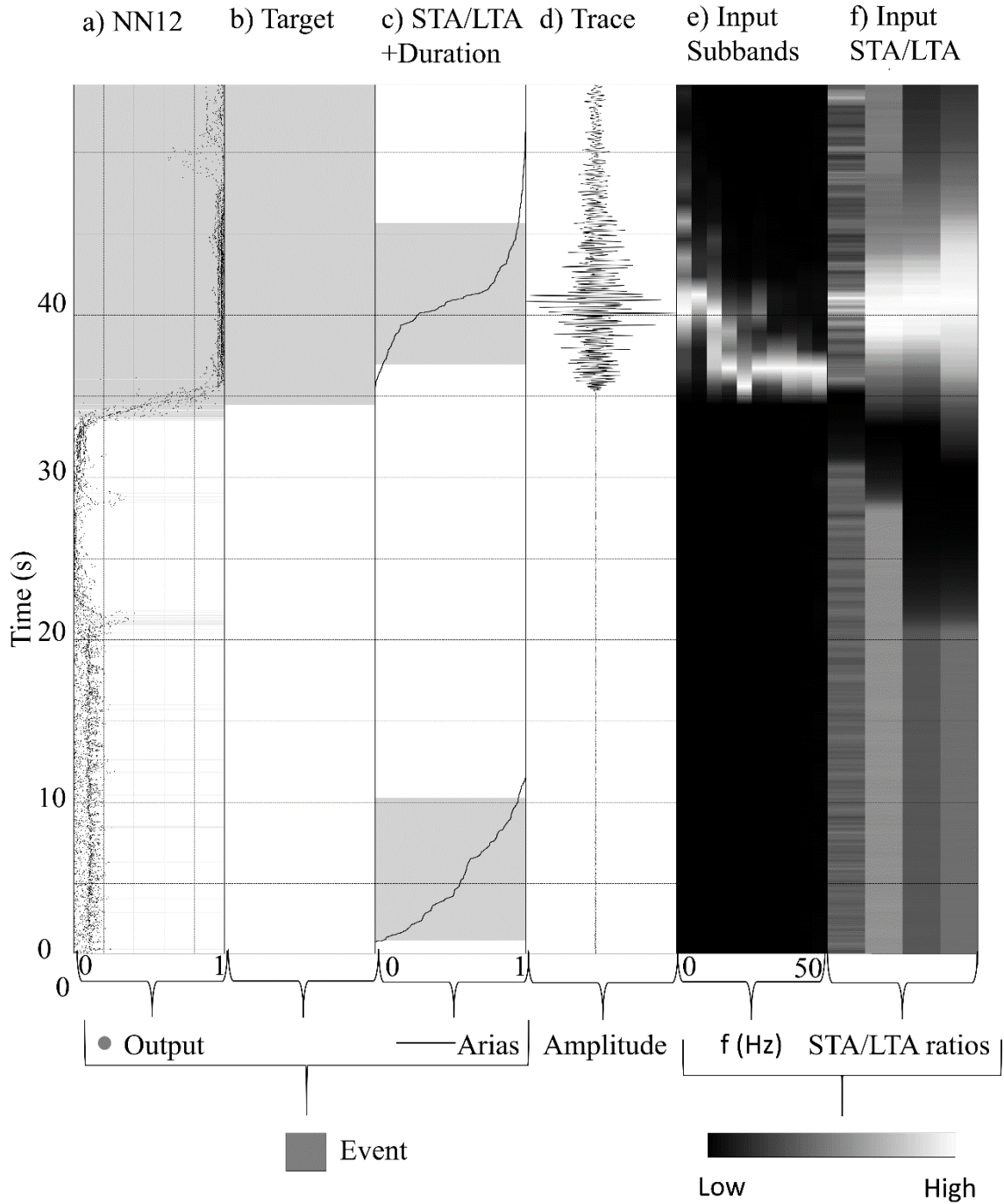

**Figure 4.** Comparison of classification results for the two detection methods together with input and target functions for the Garsthuizen event recorded on station G261 (see figure 1 for station location).(a) Results of the NN12 classification. The small dots represent classification output per time sample for the 20 realizations (i.e. 20 fold). The classification threshold was set at 0.5, above which a sample is identified as corresponding to an event and is marked gray. (b) The target function defined by the manual interpreter, (c) Results of the STA/LTA detection method combined with duration for the reference ratio STA/LTA=0.5s/10s. Here the black line represents the Arias intensity and gray marks the event corresponding to the definition of the significant duration. (d) The original input waveform. (e) The 10 frequency subband graphs ranging from 0 to 50 Hz with 5 Hz bandwidth, used as input attributes.(f) The four STA/LTAs ratio graphs used as input attributes.

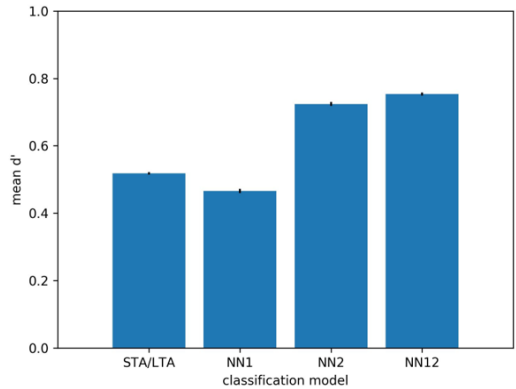
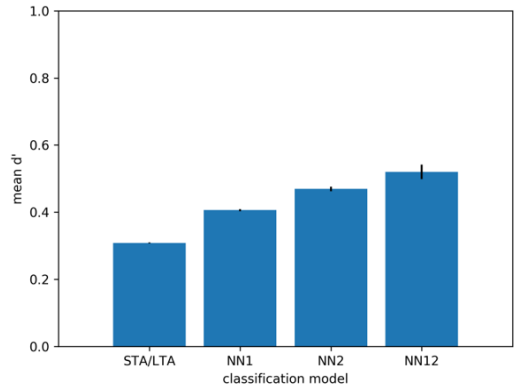

**Figure 5.** Performance of the seismic event detection methods. (a) Result for the neural network for one seismic event (Wagenborgen event). (b) Results of the neural network for all three seismic events.

In Figure 6a and b the mean ROC curve with its variance of the 20 folds trained NN12 networks are shown for respectively 1 event and 3 events. The mean sensitivity index d0 is indicated with a star. Figure 7 shows the performance of NN12 for 3 events as compared to STA/LTA with equal false positive rates, p = 0:00. This Figure shows that the neural network has a 65% increased accuracy in predicting seismic events as compared to the STA/LTA method.

In Figure 8 the performance of the STA/LTA and NN12 detection methods are plotted in a cross-plot for the three separate seismic events. In Figure 8a, b and c, each scatter point represents the cross-plot of d0 of the two detection methods for a single station. The Figure shows that the NN12 has an overall better performance than the STA/LTA method for all three seismic events. This is also indicated by the total mean of the d' (black dot), which falls well above the fictive boundary of equal performance (black dashed line). Additionally the performance of the two detection methods is compared for each event in Figure 9. This Figure shows histograms of the mean d0 per event and illustrates that for each event the neural network is dominated by relatively higher d0 values compared to the STA/LTA method.

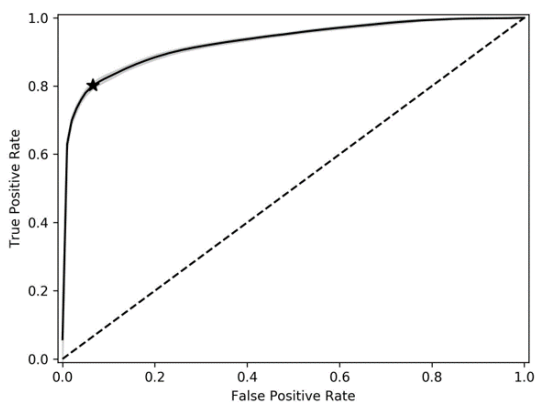
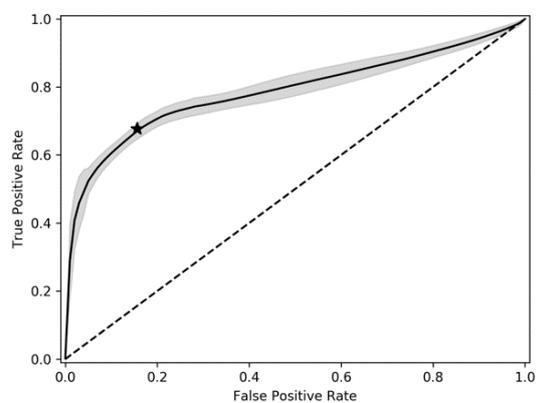

**Figure 6.** Mean ROC curve with its variance (gray) of the trained NN12 network on 3 seismic events (20 fold) and the mean sensitivity index (star). The dashed line indicates the boundary where the true positive rate equals the false rate. a) The result of the neural network trained on one event (Wagenborgen event). b) The result for the neural network trained on all three events.

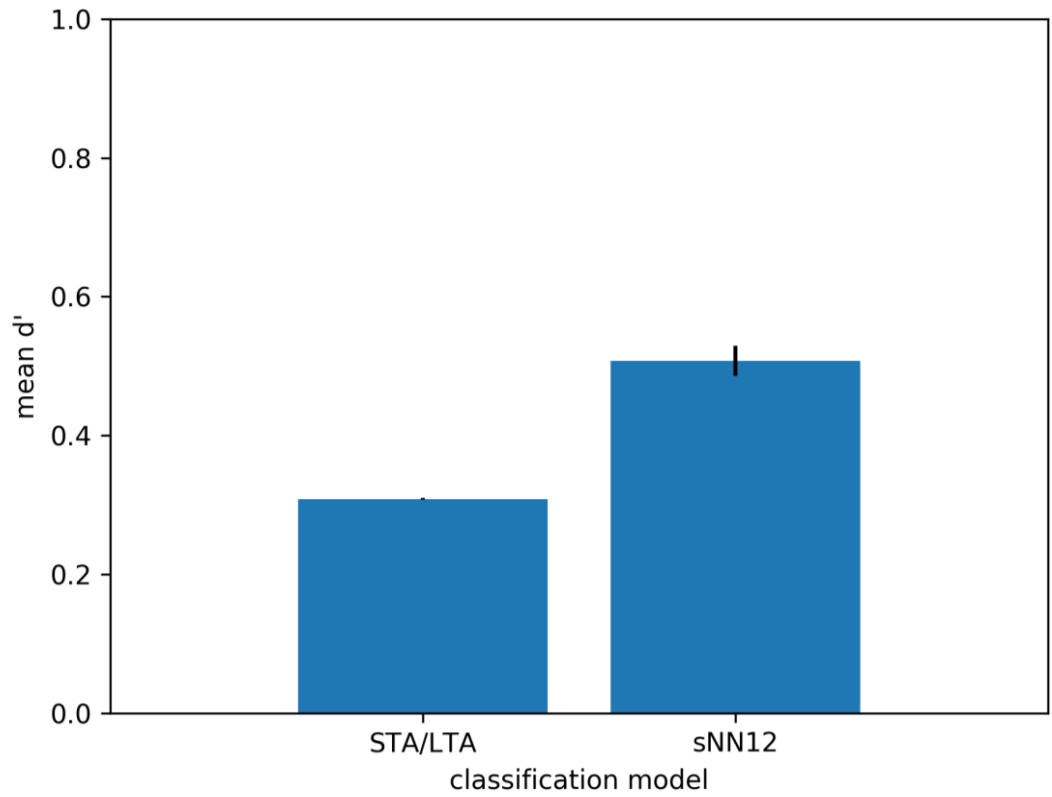

**Figure 7.** Bar plot of the performance of the STA/LTA and NN12 detection method on 3 seismic events, expressed by mean d0 with equal false positive rates.

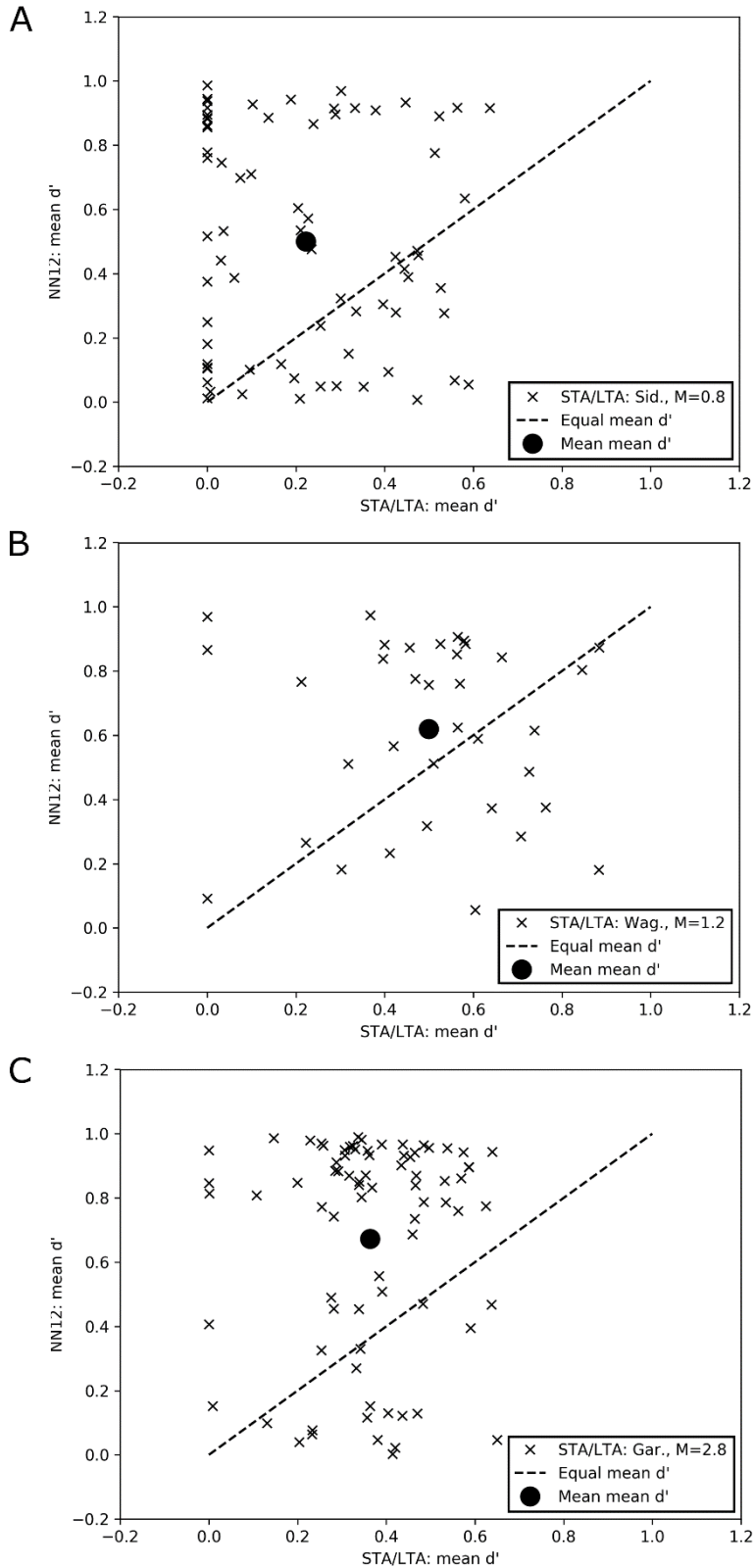

**Figure 8.** Cross-plot of the performance of the NN12 trained on 3 events against the STA/LTA detection methods expressed by d0 for each separate event. a) The Siddeburen event. b) The Wagenborgen event. c) The Garsthuizen event. The dashed line indicates the boundary where the mean d0 of the two detection methods are equal. The mean of all scatter points is marked with the black dot.

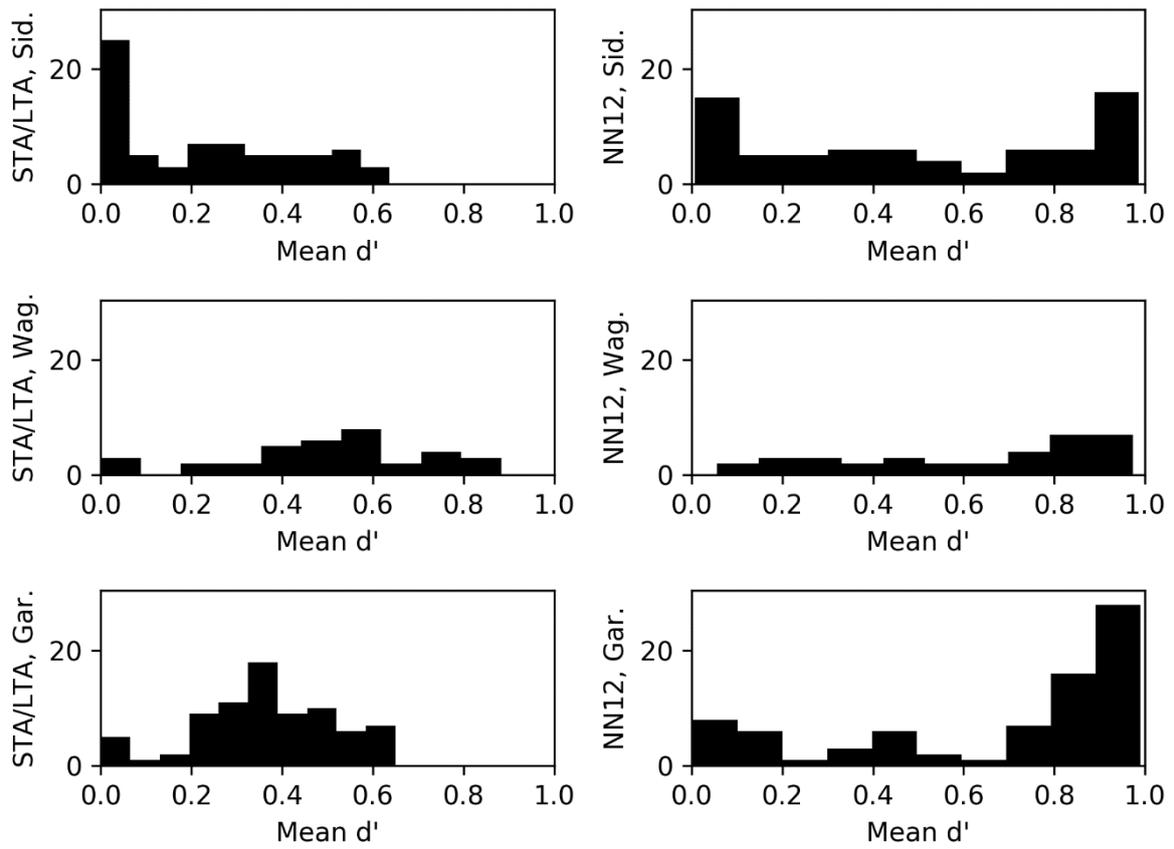

**Figure 9.** Histograms of the performance expressed by mean d0 for the three different events. The STA/LTA and NN12 outcomes are plotted, respectively, in the left and the right column.

**Discussion**

The analyses and results of the reference STA/LTA method, and the NN1, NN2 and NN12 methods demonstrate a favorable trend in neural network performance. The relative and absolute number of true positive event-onset detections and duration, with our employed neural networks is superior to the reference STA/LTA method (see Figures 5 and 7). One should note that both the event-onset times and event duration are conventionally not determined automatically from the reference STA/LTA method. A second refined picking algorithm may be conducted as well, such as an AR-AIC picker, after which a human interpreter makes a refined pick of the event onset time and duration. We foresee that the neural network used in this study can reduce this to a two-step procedure. Furthermore manual event P- and S-phase picking could be optimized by automatic detection with ANN. The significant duration on STA/LTA output can introduce additional false positive detections because the duration is extended by two time windows: 5 s before trigger on and 10 s after trigger off. This is the main reason why in standard earthquake analysis a human interpreter is involved to evaluate detected events and reject false positives, before continuing interpretation. The ANN has the strong advantage of assisting a human interpreter by limiting the number of false positives, while already providing additional information, such as duration indication, which is an important attribute especially for GFF relating building damage to seismic intensity.

The most significant achievement of the ANN is its overall improved performance for a large amount of station recordings with a significant variety in signal to noise ratios. The neural network has undeniably detected events where the reference STA/LTA method did not, and this is a workable result on which to continue expanding the neural network methodology. Since we trained the ANN on a diverse dataset, we expect that the ANN can cope with detecting other events in the GGF as well. If performance is insufficient, the ANN can be trained with more data. Computation time for training can be significantly reduced by using a GPU cluster instead of a CPU cluster, the latter being used for this study. Successive application of the neural network on new data is near instantaneous. The approach presented here lends itself to be extended for automated localization of earthquakes on (semi) real-time basis. Moreover, additional neural network training focusing on distinguishing p- and s-phases, can potentially further improve the accuracy of automated (semi) real-time calculations of earthquake arrival times and locations.

The ANN was trained and tested on 1) the single dataset of event Wagenborgen only and 2) the three datasets of events Siddeburen, Wagenborgen and Garsthuizen combined, to analyze the performance of the ANN on an increasing number and diversity of event recordings. As Figures 5, 6 and 8 indicate, the change in performance is significant, but not trivial. In Figures 5b and 6b an apparent overall performance decrease in mean d' is observed for the three events combined, but a more favorable trend in performance increase is observed with ANN complexity from STA/LTA to NN1, NN2 and NN12. We explain this overall performance decrease by the impact of more diverse event recordings and noise types. In the ANN training of only the Wagenborgen event, a fairly small and homogeneous dataset is encountered which is well recognized and classified. When confronted with the much larger Siddeburen and Garsthuizen event datasets which are of different source characteristic and are more heterogeneous in station types (geophones, broadband stations) and offset ranges, the ANN initially has a harder time to detect events than the Wagenborgen only case. But the positive outlook of using more and larger event datasets is seen in Figures 5 and 8; including more attributes of the larger datasets in the increasingly complex ANN's NN1, NN2 and NN12 not only outperforms the STA/LTA method, but also the NN performance increase with complexity is steeper in Figure 5b.

Figures 8 and 9 show how using more data in the events overall pulls the mean(mean d') to higher d' levels for the NN at the expense of the STA/LTA method. This is a well known effect in conventional supervised neural networks: they initially show a performance decrease when seeing deviating and heterogeneous labeled examples in the training, but as more labeled examples are used as input, they quickly recover and improve classification at an exponential rate. Hence, we expect the inclusion of much more event datasets to improve the ANN performance substantially. Currently, only a single channel analysis is performed, making the current method essentially a 1-D approach. For this approach the neural network method can reduce the amount of false-positive event datasets, thereby limiting data communication. Recalling the first step of seismic event detection, this single channel or single station approach is in line with the goal of fast and reliable event detection for early-warning systems and traffic light systems. This is especially relevant for large networks, where a reduction of data transfer can save energy consumption and extend battery life. Networks such as deployed at the Groningen site can benefit from such an approach, especially when it is further expanding with additional monitoring instrumentation, such as distributed acoustic sensing systems having a very high spatio-temporal sampling. When multiple stations in a network are simultaneously analyzed by a single neural network or a chain of those, multichannel patterns can be recognized, which can be accommodated in a decision support system.

**Conclusions**
This study shows that a neural network based on a combination of multiple STA/LTA sets and mean power spectral densities as input, can be trained to perform well with respect to its capability of estimating both the onset and duration of seismic events at the Groningen Gas Field. This means that the considered input sets contain enough information, are complementary and reinforcing, and the network structure is flexible enough to train a successful seismic event detector. We furthermore showed that for all three events ranging from magnitude of 2.8 down to 0.8, an increased performance is reached with the neural network detection method.

**Data and Resources**
The seismic event data used in this study was obtained from the KNMI public data portal which can be accessed at http://rdsa.knmi.nl/dataportal/ (last accessed on January 17, 2020).